\documentclass[twocolumn]{aastex63}
\usepackage{natbib}
\usepackage{hyperref}
\newcommand{\uat}[2]{\href{http://astrothesaurus.org/uat/#2}{#1 (#2)}}

\newcommand{\longname}{[VV98] J010250.2-220929}
\newcommand{\shortname}{QSO 0102$-$2209}
\newcommand{\HST}{\textit{HST}}
\newcommand{\kms}{\mbox{km\ s${^{-1}}$}}

\graphicspath{{./}{figures/}}

\received{2019 July 17}
\revised{2019 September 14}
\accepted{2019 September 18}
\published{2019 October 10}
\submitjournal{ApJL}

\shorttitle{Cooler Gas in a Filament}
\shortauthors{Connor et al.}
 
\begin{document}

\title{COS Observations of the Cosmic Web: A Search for the Cooler Components of a Hot, X-ray Identified Filament}

\correspondingauthor{Thomas Connor}
\email{tconnor@carnegiescience.edu}

\author[0000-0002-7898-7664]{Thomas Connor}
\affil{The Observatories of the Carnegie Institution for Science, 813 Santa Barbara Street, Pasadena, CA 91101, USA}

\author[0000-0001-7869-2551]{Fakhri S. Zahedy}
\affil{Department of Astronomy \& Astrophysics, The University of Chicago, Chicago, IL 60637, USA}
\affil{The Observatories of the Carnegie Institution for Science, 813 Santa Barbara St., Pasadena, CA 91101, USA}

\author[0000-0001-8813-4182]{Hsiao-Wen Chen}
\affil{Department of Astronomy \& Astrophysics, The University of Chicago, Chicago, IL 60637, USA}
\affil{Kavli Institute for Cosmological Physics, The University of Chicago, Chicago, IL 60637, USA}

\author[0000-0003-4063-5126]{Thomas J. Cooper}
\affil{The Observatories of the Carnegie Institution for Science, 813 Santa Barbara St., Pasadena, CA 91101, USA}

\author[0000-0003-2083-5569]{John S. Mulchaey}
\affiliation{The Observatories of the Carnegie Institution for Science, 813 Santa Barbara St., Pasadena, CA 91101, USA}

\author[0000-0001-8121-0234]{Alexey Vikhlinin}
\affil{Harvard-Smithsonian Center for Astrophysics, 60 Garden Street, Cambridge, MA 02138, USA}

\begin{abstract}

In the local universe, a large fraction of the baryon content is believed to exist as diffuse gas in filaments. While this gas is directly observable in X-ray emission around clusters of galaxies, it is primarily studied through its UV absorption. Recently, X-ray observations of large-scale filaments connecting to the cosmic web around the nearby ($z=0.05584$) cluster Abell 133 were reported. One of these filaments is intersected by the sightline to quasar [VV98] J010250.2$-$220929, allowing for a first-ever census of cold, cool, and warm gas in a filament of the cosmic web where hot gas has been seen in X-ray emission. Here, we present UV observations with the Cosmic Origins Spectrograph and optical observations with the Magellan Echellette spectrograph of [VV98] J010250.2$-$220929. We find no evidence of cold, cool, or warm gas associated with the filament. In particular, we set a $2\sigma$ upper limit on Ly$\alpha$ absorption of $\log(N_{\rm H\,I} / \textrm{cm}^{-2}) < 13.7$, assuming a Doppler parameter of $b=20\,\textrm{km}\,\textrm{s}^{-1}$. As this sightline is ${\sim}1100\,\textrm{pkpc}$ ($0.7R_\textrm{vir}$) from the center of Abell 133, we suggest that all gas in the filament is hot at this location, or that any warm, cool, or cold components are small and clumpy. A broader census of this system -- combining more UV sightlines, deeper X-ray observations, and a larger redshift catalog of cluster members -- is needed to better understand the roles of filaments around clusters.
\end{abstract}

\keywords{
\uat{Intergalactic medium}{813};
\uat{Cosmic web}{330};
\uat{Warm-hot intergalactic medium}{1786};
\uat{Intergalactic medium phases}{814};
\uat{Cool intergalactic medium}{303};
\uat{Hot intergalactic medium}{751};
\uat{Intracluster medium}{858};
\uat{Galaxy clusters}{584};
\uat{Ultraviolet astronomy}{1786};
\uat{Quasar absorption line spectroscopy}{1317}}

\section{Introduction} \label{sec:intro}

At its largest scales, the universe is a vast cosmic web of filamentary structure \citep[e.g.,][]{1996Natur.380..603B}. Simulations have shown that these filaments, while taking up less than 10\% of the volume of the universe hold ${\sim}40\%$ of the universe's mass \citep{2010MNRAS.408.2163A}.  Although easily traced by the locations of galaxies \citep[e.g.,][]{1986ApJ...302L...1D}, the baryonic component of this web is primarily diffuse gas \citep[e.g.,][]{2001ApJ...552..473D}. Quantifying the conditions of this gas -- the distributions of temperature and density -- is needed to balance the universal baryon budget to ensure the accuracy of cosmological models \citep{2012ApJ...759...23S}.

 \begin{figure*}
 \begin{center}
\includegraphics{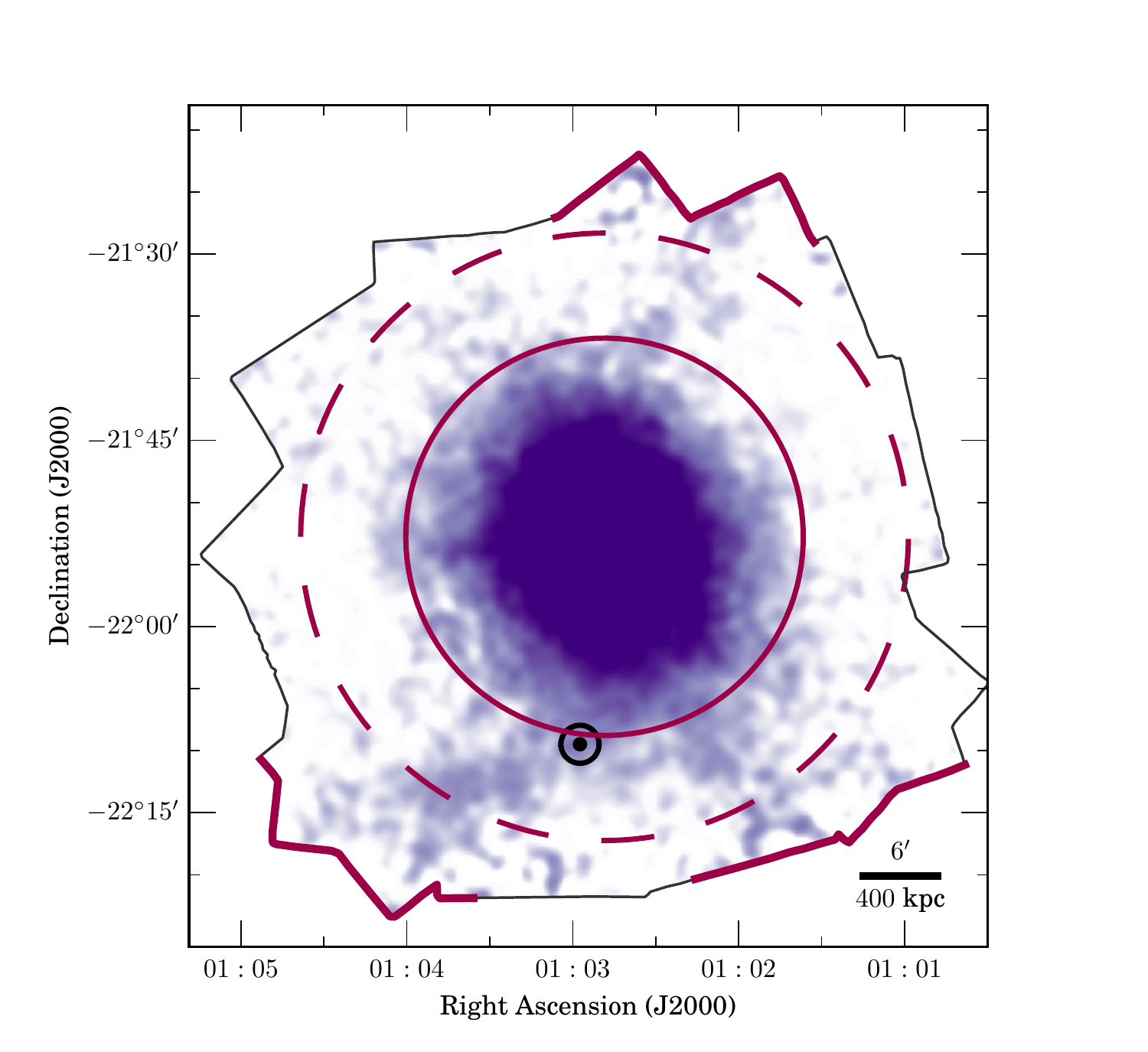}
\end{center}
\caption{Position of \shortname\ relative to the cluster Abell 133. Smoothed X-ray emission from \citet{2013HEAD...1340101V} is shown in purple, while the solid and dashed circles trace $R_{500}$ and $R_{200}$, respectively. The position of the quasar is marked by the dotted circle just south of the $R_{500}$ radius. The extents of the \textit{Chandra} observations are shown for reference; the border of these observations is highlighted in red at the positions of the filaments identified by \citet{2013HEAD...1340101V}.}\label{fig:QSO_position}
\end{figure*}

Due to the low density of filamentary gas, detections of direct emission have often proven unsuccessful \citep[e.g.,][]{1995A&A...302L...9B}. Therefore, the dominant method for studying the gas contents of filaments \citep[e.g.,][]{2016MNRAS.457.4236B} is through absorption studies \citep[see review of this technique by][]{2017ASSL..434..291C}, which require chance superpositions of bright background sources. Enabled by the tremendous power of the \textit{Hubble Space Telescope} (HST), diffuse gas has been observed at scales of the circumgalactic medium (CGM) around individual galaxies \citep[see review by][]{2017ARA&A..55..389T} out to intercluster filaments \citep{2016MNRAS.455.2662T}. In particular, cool gas ($T \sim 10^4 - 10^5\ \textrm{K}$) is best traced through narrow neutral hydrogen (\ion{H}{1}) absorption \citep[e.g.,][]{2019MNRAS.484.2257Z}.

\begin{figure*}
\begin{center}
\includegraphics{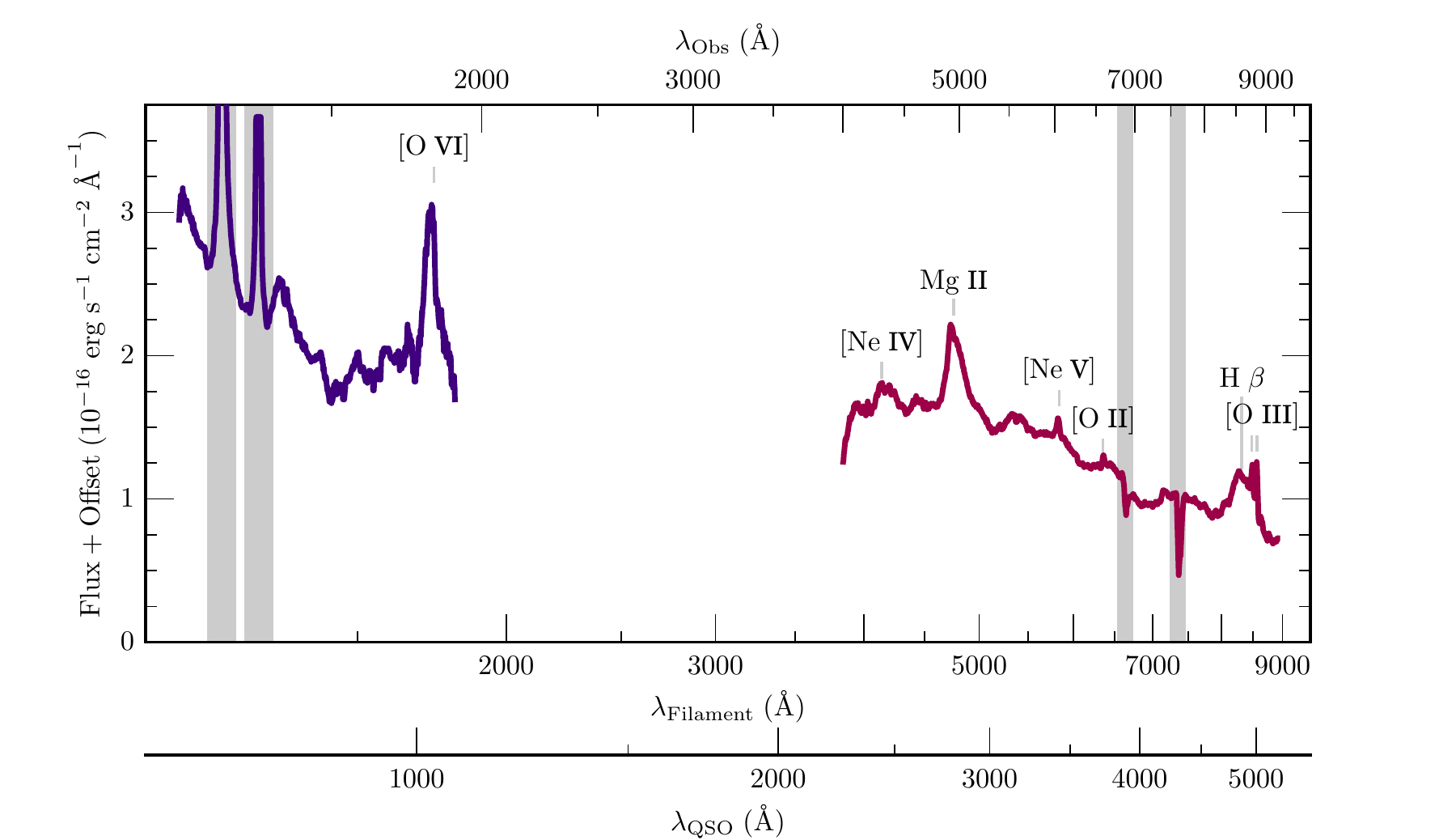}
\end{center}
\caption{UV and optical spectrum of \shortname\ from COS (purple) and MagE (maroon). Prominent emission features of the quasar are labeled, and regions of terrestrial contamination are shaded. For display purposes, both spectra have been smoothed and low signal-to-noise regions have been removed. As a reference, the wavelength scales as observed (top), in the rest-frame of the filament (upper bottom), and in the rest-frame of the quasar (lower bottom) are given.}\label{fig:full_spectrum}
\end{figure*}

Alternatively, at the nodes of filaments, the gas in groups and clusters of galaxies is easily observed and studied in X-ray emission \citep[e.g.,][]{2014ApJ...794...48C}. There have been some successes in pushing outward from clusters to see the X-ray emission from filaments, notably from merging clusters (e.g., \citealt{2008A&A...482L..29W}; merging increases the temperature and density of a filament, making it X-ray brighter), and around Abell 2744 \citep{2015Natur.528..105E}, which has extreme amounts of substructure \citep{2017MNRAS.467.2913S}. In an extremely deep observation of Abell 133 (2.4 Ms), \citet{2013HEAD...1340101V} reported seeing extended X-ray emission from filaments that extend beyond the cluster's virial radius, as shown in Figure \ref{fig:QSO_position}. A follow-up analysis and redshift survey by \citet{2018ApJ...867...25C}  found that these structures are traced by the galaxy population and align with the locations of large-scale structure filaments.

Included in that survey is a potentially interesting target: \longname\ (also known as XRS 0102$-$2209, hereafter \shortname), which was identified by \citet{1998MNRAS.299.1047W} as an active galactic nucleus behind Abell 133. In archival observations with the {\it Galaxy Evolution Explorer} \citep{2007ApJS..173..682M} the quasar has an FUV brightness of  $20.77 \pm 0.26$\ mag and an NUV brightness of $20.42 \pm 0.16$\ mag. \shortname\ is projected along one of the filaments identified by \citet{2013HEAD...1340101V} and at ${\sim}1100\,\textrm{pkpc}$  ($0.7R_\textrm{vir}$) from the center of Abell 133, it is outside $R_{500}$, as is shown in Figure \ref{fig:QSO_position}. \shortname\ therefore provides the perfect opportunity to obtain the first comprehensive baryonic census along a filament of the cosmic web.

Previous works on quasar absorption associated with clusters have focused on the nearby Virgo \citep{2012ApJ...754...84Y} and Coma clusters \citep{2017ApJ...839..117Y}. These authors reported that cool gas traced by Ly$\alpha$ mostly avoids areas of X-ray emission around clusters. Likewise, \citet{2018MNRAS.475.2067B} found no warm gas (as traced by \ion{O}{6} absorption) within $\pm 2000\,\textrm{km}\,\textrm{s}^{-1}$ along three quasar sightlines intersecting five cluster systems from $\sim 0.2 - 2.5 R_{\rm Vir}$, but did see narrow Ly$\alpha$ absorption just outside $r_{500}$ of one of a pair of merging clusters. Further complicating this picture is the work of \citet{2017ApJ...846L...8M}, who observed three quasar sightlines covering $(1.6 - 4.7) r_{500}$ of three clusters and reported that Sunyaev--Zel'dovich effect-selected clusters were more rich in cool gas than X-ray detected clusters. In previous studies of cluster outskirts \citep{2019MNRAS.488.5327P} and filaments \citep{2016MNRAS.455.2662T}, \ion{H}{1} and metals have been seen with column densities of $\log(N_{\rm H\,I}/\textrm{cm}^{-2}){\sim}14$ and Doppler parameters $b{\sim}20\,\textrm{km}\,\textrm{s}^{-1}$, although with covering fractions of the order 0.2 \citep{2019MNRAS.490.4292B} to 1.0 \citep{2016MNRAS.455.2662T}, respectively. As almost all further work on this topic has been theoretical \citep{2015MNRAS.453.4051E,2019MNRAS.490.4292B}, observations of \shortname\ provide new insights into the nature of warm and cool gas around clusters.

In this Letter, we describe UV and optical spectroscopic observations of \shortname\ to quantify the amount of absorbing gas in the X-ray identified filament. Our observations are described in Section \ref{sec:obs}, the results are given in Section \ref{sec:results}, and finally we discuss these results in the broader context in Section \ref{sec:discussion}. Based on \citet{2018ApJ...867...25C}, we adopt a redshift for Abell 133 of $z=0.05584$, and we assume a flat $\Lambda$CDM cosmology with $\Omega_M = 0.3$, and $\textrm{H}_0 = 70\ \textrm{km}\ \textrm{s}^{-1}\ \textrm{Mpc}^{-1}$. We follow the naming convention of \citet{2017ARA&A..55..389T} when referencing gas temperatures: cold ($T<10^4\,\textrm{K}$), cool ($10^4\,\textrm{K}<T<10^5\,\textrm{K}$), warm ($10^5\,\textrm{K}<T<10^6\,\textrm{K}$), and hot ($T>10^6\,\textrm{K}$). To contrast with the hot X-ray detected gas \citep{2013HEAD...1340101V}, we also adopt the term ``cooler gas'' to refer to all gas of temperature $\textrm{T}<10^{6}\,\textrm{K}$ potentially detected in absorption.


 \section{Observations} \label{sec:obs}
 \subsection{HST-COS}
The quasar \shortname\ was observed with the Cosmic Origins Spectrograph \citep[COS;][]{2012ApJ...744...60G} on \HST\ as part of program GO-15198 (PID: 15198, PI: Connor); the spectrum is shown in Figure \ref{fig:full_spectrum}. Observations were conducted over two visits: two orbits on 2018 December 21 and three orbits on 2018 December 29. We used the primary source aperture and the G140L grating with a central wavelength of 1105 \AA. This configuration provided a spectral resolving power $R \sim 1000$ with a dispersion of $80.3\ \textrm{m\AA}\ \textrm{pix}^{-1}$ and resulted in a broad UV wavelength coverage of ${\approx}1100 - 2200$ \AA. Although \ion{O}{6} cannot be observed at this redshift with COS, our spectrum included Ly$\alpha$ and prominent low-, intermediate-, and high-ionization metal absorption transitions at the cluster redshift, including  \ion{C}{2} $\lambda1334$ \ion{Si}{2} $\lambda1260$, \ion{Si}{3} $\lambda1206$, and the \ion{Si}{4} $\lambda \lambda1393,1402$ doublet. The total integration time over the two visits was 13,094 s, which was
divided into 10 individual exposures of roughly equal durations. During the second visit, the \HST\ Fine Guidance Sensors did not acquire the guide stars until approximately 45 minutes after the exposures were scheduled to start, and the early parts of these observations were therefore taken under gyro pointing control. A close examination of the data revealed that this incident had no significant impact on the data quality.
 
Individual spectral images were initially processed using the automatic COS calibration pipeline, \textsc{Calcos}, which produced a series of data products including an extracted one-dimensional  spectrum for each science exposure.  These pipeline-calibrated data products were retrieved from the {\it HST} archive. The extracted one-dimensional spectra from individual exposures were co-added to form a single combined spectrum, using inverse variance weighting based on the mean variance of the data at $1230-1250$ \AA, chosen for a lack of strong absorption features. Finally, the combined spectrum was continuum normalized using a low-order polynomial fit to spectral regions free of strong absorption features. The final continuum-normalized spectrum has a typical signal-to-noise ratio of ${\rm S}/{\rm N} \approx 8-15$ per resolution element of ${\approx} 300$ \kms\ at $\lambda \lesssim 1700$ \AA, where relevant absorption transitions are located.

 \subsection{MagE}
We observed \shortname\ with the $1\farcs0$ slit on the Magellan Echellette spectrograph \citep[MagE;][]{2008SPIE.7014E..54M} on 2017 June 18. MagE provides spectral coverage from ${\sim}3100$\ \AA\ to ${\sim} 1\ \mu\textrm{m}$ micron with a spectral resolving power $R=4100$. Weather conditions were clear and calm, with seeing in the range of $0\farcs8 - 1\farcs3$. We took nine 900 s exposures of \shortname, and observed the standard star LTT 1020 for four 90 s exposures in the middle of our observations. To ensure accurate wavelength calibration, we took ThAr hollow cathode tube exposures bracketing each set of science observations.

Observations were reduced with the MAGE Spectral Extractor \citep[MASE;][]{2009PASP..121.1409B}. MASE handles the reduction and calibration of point sources from bias subtraction to flux calibration, and can achieve wavelength accuracy to ${\sim}5\ \textrm{km}\ \textrm{s}^{-1}$ and relative flux calibration to ${\sim}10\%$ \citep{2009PASP..121.1409B}. As there were issues with flat-fielding the bluest five orders of our observations, we trim our MagE spectrum to only include data redward of 4000 \AA. Note that none of the optical features we are concerned about are lost by this trimming.  The spectrum is shown in Figure \ref{fig:full_spectrum}. Using the ASERA toolkit \citep{2013A&C.....3...65Y} to identify the peaks of broad-line emission in \ion{Mg}{2} and H$\beta$, we compute a quasar redshift of $z = 0.7548$, a refinement from the value reported by \citeauthor{1998MNRAS.299.1047W} (\citeyear{1998MNRAS.299.1047W}; $z=0.766$).

\section{Results} \label{sec:results}

Analysis of the hot gas content of the filaments as revealed by X-ray observations will be presented in a future paper (Vikhlinin et al., in preparation); here, we summarize those results, as described by \citet{2013HEAD...1340101V}. The three filaments are fit by a gas with temperature $T_X\approx2.0\pm0.3\,\textrm{keV}$ (${\sim}2.3\pm 0.3\times10^{7}\,\textrm{K}$). Assuming the filamentary emission is fit by a $\beta$-model with $\beta=2/3$ and $r_c\approx500\ \textrm{kpc}$, the integrated column density is of the order $\log(N_\textrm{H} / \textrm{cm}^{-2}) \approx 20 - 21$ for a sightline passing through the center of the filament.

To constrain the amount of cooler gas present in this sightline, we turn to our UV and optical spectra of the QSO. The QSO spectra described in Section 2 provide limiting constraints for the properties of cold, cool, and warm absorbing gas in the intervening filament, as traced by absorption of neutral hydrogen and ionized metal species expected from both photoionized and collisionally ionized warm gas. In the UV spectrum taken with COS, these species include neutral hydrogen \ion{H}{1}, low-ionization metals \ion{C}{2} and  \ion{Si}{2}, intermediate-ionization metal  \ion{Si}{3}, and highly ionized \ion{Si}{4}. In addition to the species probed by our COS observations, we also looked at \ion{Ca}{2} $\lambda\lambda3934,3969$ doublet in the optical MagE spectrum, which acts as an independent tracer of neutral hydrogen and dust \citep[e.g.,][]{2006MNRAS.367..211W}.
\begin{figure*}
\begin{center}
\includegraphics{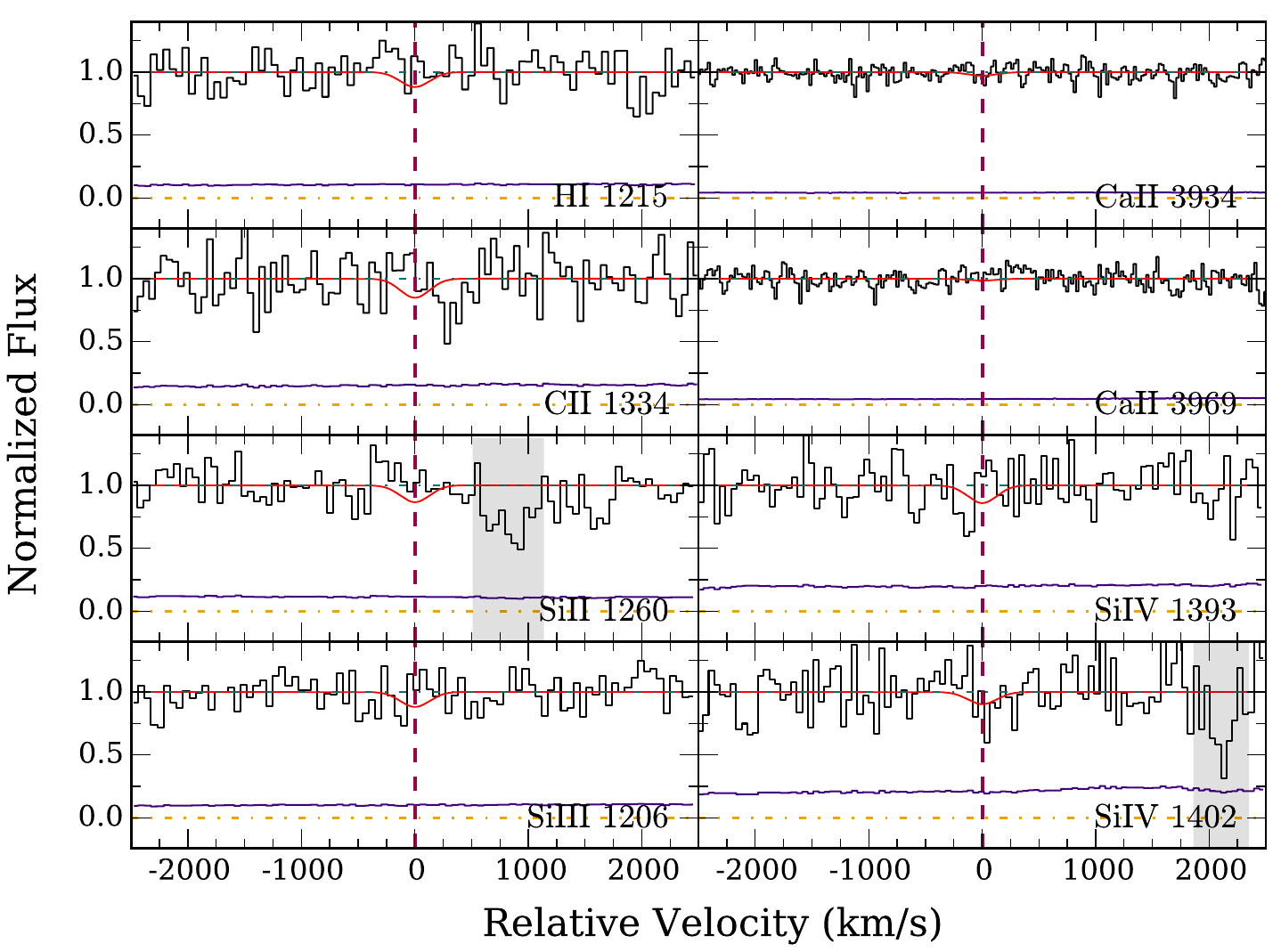}
\end{center}
\caption{Continuum-normalized spectrum for different absorption transitions along \shortname\ at $d=1090$ pkpc ($0.7R_\textrm{vir}$) from the center of Abell 133. The absorption transition is identified in the bottom right corner of each panel. Zero velocity marks the redshift of Abell 133 at $z=0.05584$. The $1\sigma$ error spectrum is included in purple above the zero-flux level. Contaminating features (Galactic \ion{C}{2} $\lambda1334$ and an unassociated intervening absorber) have been grayed out for clarity. We find no evidence of the presence of cooler gas within $\pm2500$ \kms\ from the cluster redshift, with a sensitive $2\sigma$ column density upper limit of log(\,$N_\mathrm{H\,I} / \textrm{cm}^{-2})<13.7$ for neutral hydrogen \ion{H}{1} absorption. The $2\sigma$ upper limits on these absorption transitions are plotted in red curves for a single-component Voigt profile with $b=20$ \kms.}\label{fig:COS_absorption_suite}
\end{figure*}

\begin{deluxetable}{llrr}
\tablecaption{Constraints on Absorption Properties}
\tablewidth{0pt}
\tablehead{
\colhead{Species} & \colhead{Transition} & \colhead{$W_r$\tablenotemark{a}} & \colhead{$\log(N/\textrm{cm}^{-2})$\tablenotemark{b}} \\
\colhead{} & \colhead{(\AA)} & \colhead{(m\AA)} & \colhead{} }
\startdata
\ion{H}{1} & 1215 & \textless160 & \textless13.7 (13.5, 13.5) \\
\ion{C}{2} & 1334 & \textless238 & \textless14.4 (14.2, 14.1) \\
\ion{Ca}{2} & 3934 & \textless127 & \textless12.2 (12.2, 12.2) \\
\ion{Si}{2} & 1260 & \textless177 & \textless13.4 (13.2, 13.1) \\
\ion{Si}{3} & 1206 & \textless153 & \textless13.1 (12.9, 12.9) \\
\ion{Si}{4} & 1393 & \textless231 & \textless13.7 (13.5, 13.4)
\enddata
\label{tab:AbsResults}
\tablenotetext{a}{All reported values are $2\sigma$ upper limits}
\tablenotetext{b}{Converted from the upper limit on $W_r$ assuming a single-component profile with $b=20$ \kms. Upper limits in parentheses assume $b=50$ and $100$ \kms, respectively.}
\end{deluxetable}

We searched for absorption features in the spectra of the background QSO over a line-of-sight velocity interval of $\pm2500$ \kms\ from the systemic redshift of Abell 133. This search window corresponds to approximately $\pm 3 \sigma_\mathrm{v}$, where $\sigma_\mathrm{v}$ is the observed line-of-sight velocity dispersion of Abell 133 \citep{2018ApJ...867...25C}, and is therefore sufficiently wide to include all absorption features which are likely to be physically associated with the galaxy cluster.  As shown in Figure 3, we do not find evidence of a significant Ly$\alpha$ absorption within the search window. We place a sensitive $2\sigma$ rest-frame equivalent width upper limit on Ly$\alpha$ absorption of  $W_\mathrm{r}(1215) <0.16$ \AA, integrated over a velocity range of $\pm 300$ \kms\ from the cluster redshift and calculated using the $1\sigma$ error spectrum. The chosen integration window is twice the spectral FWHM of our COS data. This equivalent width limit corresponds to a column density upper limit of  log(\,$N_\mathrm{H\,I} / \textrm{cm}^{-2})<13.7$ for a single-component Voigt profile with Doppler width of $b=20$ \kms, which is typical of \ion{H}{1} absorbers in cluster environment \citep[e.g.,][]{2016MNRAS.455.2662T,2018MNRAS.475.2067B,2019MNRAS.488.5327P}. Likewise, we do not detect any significant absorption from low-, intermediate-, or high-ionization metals within the search window. We show the results of our absorption search in Table \ref{tab:AbsResults} and Figure \ref{fig:COS_absorption_suite}; in Table \ref{tab:AbsResults}, for every ionic species we present the upper limits on the rest-frame absorption equivalent widths $W_r$ and ionic column density $\log(N)$, calculated for the strongest available transition at the $2\sigma$ level and integrated over 600 $\textrm{km}\ \textrm{s}^{-1}$ windows centered at the redshift of Abell 133. 

In Table \ref{tab:AbsResults} we also provide column density upper limits for assumed Doppler parameters of $b = 50\ \textrm{and}\ 100$ \kms. 20 \kms\ is perhaps a lower limit on the velocity broadening based on the amount of turbulence expected at the cluster outskirts \citep[e.g.,][]{2017MNRAS.469.3641I}. In contrast, to produce a broadening of $b=100\,\textrm{km}\,\textrm{s}^{-1}$ from thermal motions, \ion{H}{1} would have a temperature of $6\times10^{5}\,\textrm{K}$. Due to the relatively low spectral resolution of the G140L grating on COS (compared to the G130M grating), our resolution element is only ${\sim}300\,\textrm{km}\,\textrm{s}^{-1}$, which means we cannot resolve any differences in intrinsic line width at these levels. Considering the results of prior studies \citep{2016MNRAS.455.2662T, 2019MNRAS.488.5327P}, we restrict our discussion to the results when we assume $b=20\,\textrm{km}\,\textrm{s}^{-1}$, but the exercise shown in Table \ref{tab:AbsResults} demonstrates that the upper limits on column densities would be slightly tightened if we assume a stronger contribution from turbulence broadening for a $\sim10^{4}\,\textrm{K}$ cool gas, or if we assume a warmer gas of a few$\times 10^{5}\,\textrm{K}$.

\section{Discussion} \label{sec:discussion}

As part of our observations of \shortname\, we were unable to observe any evidence of cold, cool, or warm gas (hereafter ``cooler gas'') coincident with the hot gas reported by \citet{2013HEAD...1340101V}. We discuss two potential physical reasons for this lack of absorption features: the entirety of the cooler gas in the filament this close to the cluster has been heated or the distribution of cooler gas in the filament is nonuniform. Below, we discuss both options individually, as well as the contribution of CGM from infalling galaxies to the cosmic filaments.
 \begin{figure*}
 \begin{center}
\includegraphics{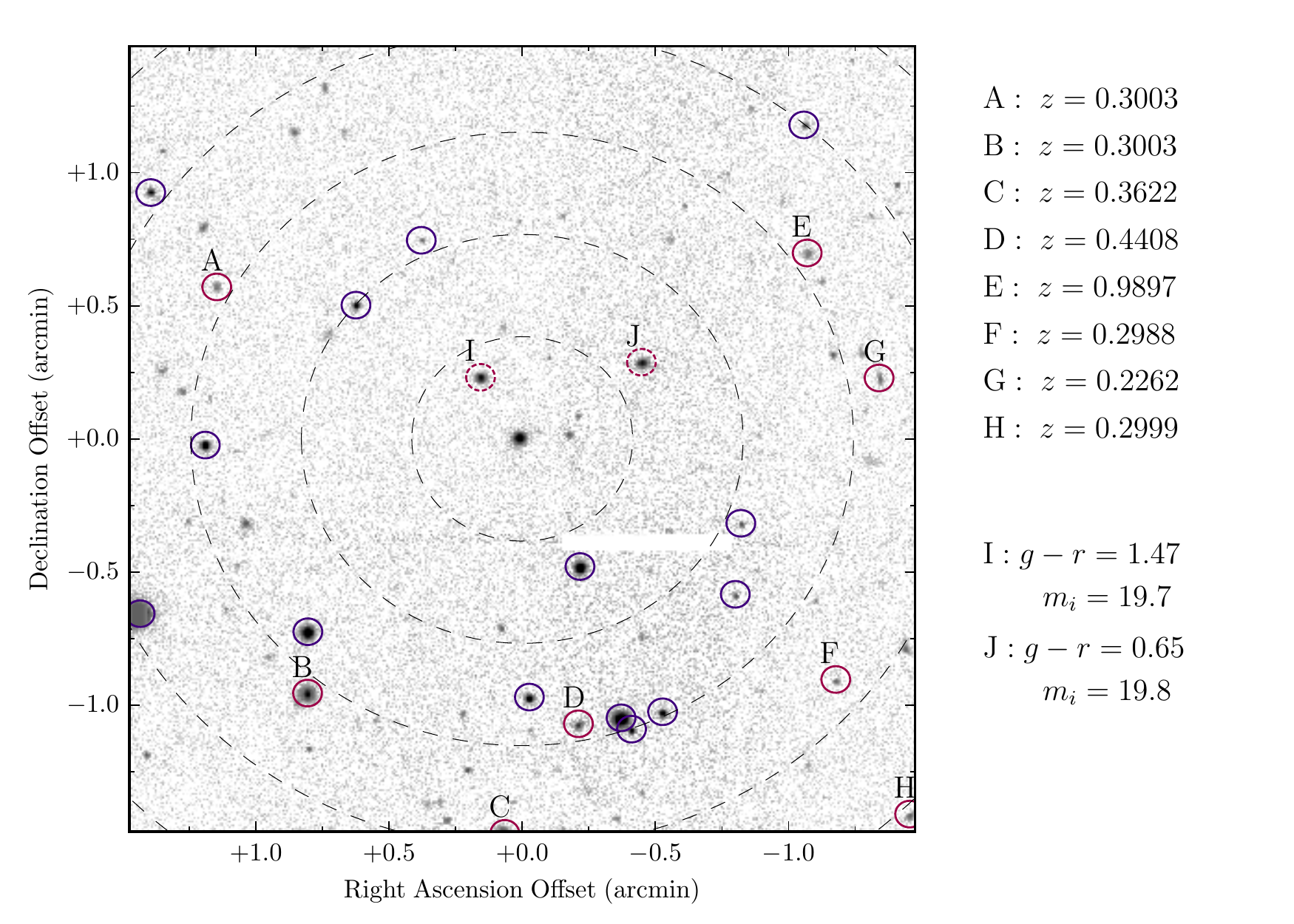}
\end{center}
\caption{$i$-band image of the field around \shortname\ from the DES \citep{2018ApJS..239...18A}. Dashed black circles denote impact parameters in multiples of 25 pkpc at the cluster redshift. Stars \citep[selected from morphology as described in][]{2018ApJS..239...18A} are indicated with purple circles, while galaxies are marked with maroon circles. Where available, redshifts from \citet{2018ApJ...867...25C} are shown to the right; for other galaxies brighter than $i < 21$, DES photometry is given. For reference, the larger black circle in Figure \ref{fig:QSO_position} has a radius of 100 pkpc.}\label{fig:DES_cutout}
\end{figure*}

\subsection{Fully Heated Filaments} \label{ssect:heated}

The most obvious potential reason for our spectra not showing signs of absorption is that the filament is fully heated this close to the cluster, such that all gas at this location is now hot. Simulations by, e.g., \citet{2018MNRAS.476.4629P} have shown that the centers of clusters are lacking in cooler gas, which begins to be seen outside the virial radius. \citet{2015MNRAS.453.4051E} report a potential link between ongoing merging and observed warm gas absorption; as Abell 133 is relatively dynamically relaxed \citep{2018ApJ...867...25C}, cool clumps should have had time to reach thermal equilibrium.

While presenting simulations of the cosmic web, \citet{2006MNRAS.370..656D} described a phenomenon of advancing accretion shocks, which caused temperatures to rise two orders of magnitude along filaments out to twice the virial radius. Behind these shocks, gas is isothermal. Similar shock heating along filaments has been reported in observations of bridges in early-stage cluster mergers \citep[e.g,.][]{2015PASJ...67...71K}. An accretion shock along the filamentary axis leaving isothermal hot gas in its wake would explain both the ability to observe the filaments in X-rays and the lack of Ly$\alpha$ absorption.

Although not as tied to a particular phenomenon, a similar result is seen in recent simulations by \citet{2019MNRAS.490.4292B}. They describe filaments of warm gas visible just outside the virial radius, but which are rare near the cluster center. In similar simulation work, \citet{2015MNRAS.453.4051E} saw that low column density absorbers trace infalling filaments, but do not survive their first infall, and that the cooler gas mass fraction is less than a few percent within the cluster virial radius. The combined picture from all the works described here is that our sightline may be located too close to the cluster center, and that any gas that was warm further out along the filament may have already been heated by the intracluster medium.

More broadly, simulations by \citet{2019MNRAS.486.3766M} show that the primary component of gas in filaments at $z=0$ is the warm-hot intergalactic medium (WHIM; $10^5\,\textrm{K}<T<10^7\,\textrm{K}$; \citealt{1999ApJ...514....1C}), which transitions to $T>10^7\,\textrm{K}$ gas closer to bound structures. As the baryon mass fraction in filaments cannot be completely recovered by X-ray emitting gas \citep{2015Natur.528..105E}, and as we set upper limits on most other diagnostics of cooler gas, we expect that there may still be a component of WHIM traced by \ion{O}{6} absorption, which is inaccessible to COS at this redshift. This aligns with other recent works (e.g., \citealt{2016MNRAS.455.2662T}; \citealt{2018Natur.558..406N}; but see also \citealt{2019ApJ...884L..31J})

\subsection{Nonuniform Distribution of Warm Gas}

One important caveat with our analysis is that we can only probe one sightline through the filament. In the case that there is cooler gas that is not uniformly distributed, our nondetection of Ly$\alpha$ may be related more to the relative positions of Earth and \shortname\ than to the overall conditions of filamentary gas.

Simulations have long seen an increase in gas clumping at the outskirts of clusters \citep[e.g.,][]{2011ApJ...731L..10N}. Similarly, observational work by, e.g., \citet{2014MNRAS.439.1796W} has found that the gas at the outskirts of clusters is clumpy and multiphase. \citet{2014MNRAS.439.1796W} also connect the increasing clumpiness with the decreasing impact of nongravitational feedback, which will also hold in filaments. 

The conditions of the filaments reported by \citet{2013HEAD...1340101V} are similar to that of a galaxy group ($T_X \sim 2\ \textrm{keV}$), so we also consider two works related to gas in group environments. \citet{2017ApJ...844...23P} put forth an ``onion-skin'' model for ionization states of CGM, with the harsh group environment ionizing gas from the outside in. Assuming a similar effect is happening to cooler gas clumps, we would expect them to be relatively small this close to the cluster center. \citet{2019ApJS..240...15S} report that in groups warm gas is patchy and has a small covering fraction (less than 10\%). If we assume the cooler gas covering fraction is similar to that of the intracluster medium of Abell 133, we would expect it to be between that seen for the Virgo cluster (which is approximately half the mass of Abell 133) and the Coma cluster (approximately three times the mass of Abell 133). Inside the virial radius, \citet{2017ApJ...839..117Y} report covering fractions for $\log(N_{\rm H\,I} / \textrm{cm}^{-2})\geq13.8$ absorbers is ${0.25}_{-0.13}^{+0.24}$ and ${0.09}_{-0.05}^{+0.13}$ for Coma and Virgo, respectively.

\vspace{1cm}
\subsection{CGM of Filament Galaxies}

We also use our one sightline to place limits on the amount of CGM that galaxies in filaments can retain. In clusters, works by \citet{2013ApJ...772L..29Y}, on the Virgo cluster, and \citet{2018MNRAS.475.2067B}, on X-ray selected clusters, have shown that the covering fraction of CGM for cluster galaxies is greatly reduced compared to field galaxies. To explain this phenomenon, previous observational work has shown evidence for preprocessing of galaxies along filaments (e.g., \citealt{2019MNRAS.484....2S} and references within), which theoretical works often attribute in some way to infalling galaxies losing their supply of gas \citep[e.g.,][]{2019OJAp....2E...6A}. These observations can provide evidence of this effect, as strict limits on absorbing gas are observed, if any filament-associated galaxies are near our sightline.

We show the area around \shortname\ using imaging from the Dark Energy Survey \citep[DES;][]{2018ApJS..239...18A} in Figure \ref{fig:DES_cutout}. Excluding both stars and galaxies with spectroscopic redshifts from \citet{2018ApJ...867...25C}, of which none are cluster members, only two other galaxies brighter than $m_i < 21$ remain within ${\sim}100\ \textrm{pkpc}$ at the cluster redshift. The first (labeled ``I'' in Figure \ref{fig:DES_cutout}) is too red to be at the redshift of Abell 133 \citep{2018ApJ...867...25C}. Galaxy ``J,'' however, has the photometry to be on the red sequence. With a distance modulus of $M-m\approx-37$, this galaxy would be ${\lesssim}0.1 L\textrm{*}$ \citep{2017ApJ...848...37C}; comparing this to work by \citet{2015MNRAS.449.3263J}, we would expect $\log(N_{\rm H\,I} / \textrm{cm}^{-2})\approx17$ for a similar field galaxy. If follow-up spectroscopy shows this galaxy is at the cluster redshift, the lack of observed absorption provides further confirmation that infalling galaxies are losing their CGM. This is the only potential cluster galaxy within 100 pkpc of \shortname\ that is brighter than $m_i < 21$.

This analysis is limited to only those galaxies near the sightline (beyond 100 pkpc, nondetections at the level presented here in this dense of an environment become uninteresting; e.g., \citealt{2013ApJ...772L..29Y}, \citealt{2019MNRAS.490.4292B}) and bright ($m_i=21$ is both roughly the limit of the redshift survey by \citealt{2018ApJ...867...25C} and, at ${\sim}0.005L\textrm{*}$, the limit studies of faint galaxies have explored; e.g., \citealt{2014ApJ...796..136B}). Relaxing either constraint reveals more cluster galaxies, which may be relevant for deeper studies of this sightline. Within 175 pkpc, there are at least four other cluster galaxies, based on the catalog of \citet{2018ApJ...867...25C}, including one with brightness $m_i=15.4$. Similarly, visible just to the west of \shortname\ in Figure \ref{fig:DES_cutout} are a pair of faint galaxies; while the southern one is far too red to be a cluster member, the northern galaxy of the pair, with brightness $m_i=22.8$ and color $g-r=0.58$, is potentially a dwarf galaxy in the filament. Deeper UV observations are needed to draw meaningful insights into either population, however.

\section{Summary}

We present UV and optical observations of \shortname, a quasar whose sightline intersects X-ray detected filaments around the galaxy cluster Abell 133. We find no evidence for cool gas ($T \leq 10^{5}\ \textrm{K}$, as traced by $\textrm{Ly}\ \alpha$), to limits of $\log(N_{\rm H\,I} / \textrm{cm}^{-2}) < 13.7$ (assuming $b=20\,\textrm{km}\,\textrm{s}^{-1}$), or of cold or warm gas to similar limits. With only one sightline, we avoid drawing any significant conclusions, but this lack of observed absorption may be indicative that cosmic filaments with observed hot gas lack cooler gas as they approach clusters. Further investigations will be needed to constrain this effect, which is important in balancing the baryon budget of the universe. In addition, despite the presence of a potential filament galaxy within ${\sim}30$\ pkpc, we do not observe any CGM.

Abell 133 provides a unique opportunity to study the roles of filaments in galaxy evolution, to quantify the baryon budget, and to understand how clusters are tied to the cosmic web. Further UV observations of quasars around this cluster, deeper X-ray observations (particularly with next-generation X-ray observatories), and a complete redshift census of nearby galaxies will all contribute to developing our understanding of these three questions. 

\vspace{2mm}

{\small T.Connor was supported by STScI/NASA award \HST-GO-15198. This Letter includes data gathered with the 6.5 m Magellan Telescopes located at Las Campanas Observatory, Chile. Based on observations made with the NASA/ESA {\it Hubble Space Telescope}, obtained at the Space Telescope Science Institute, which is operated by the Association of Universities for Research in Astronomy, Inc., under NASA contract NAS 5-26555. These observations are associated with program \#15198. Support for program \#15198 was provided by NASA through a grant from the Space Telescope Science Institute, which is operated by the Association of Universities for Research in Astronomy, Inc., under NASA contract NAS 5-26555.}
 
\vspace{5mm}
\facilities{HST (COS), Magellan:Baade (MagE), Blanco}

\software{MASE \citep{2009PASP..121.1409B}, PyFITS \citep{1999ASPC..172..483B}, ASERA \citep{2013A&C.....3...65Y}}
 
\bibliography{bibliography}

\end{document}